\documentclass[12pt]{article}
\usepackage[english]{babel}
\usepackage{amsfonts}
\usepackage{amsmath}
\usepackage{amssymb}
\usepackage{mathrsfs}
\usepackage{multicol}
\usepackage{fancybox}
\usepackage{dsfont}
\usepackage{color}
\usepackage{hyperref}
\usepackage{subfigure} 
\usepackage{graphicx} 
\usepackage{caption}
\usepackage{lineno}

\def\be{\begin{equation}}
\def\ee{\end{equation}}
\def\tl{\tilde} 
 
\def\gm{\gamma} 
\def\lm{\lambda}

\def\d'{``}
\def\i{\textrm{i}}
\newtheorem{thm}{Theorem}[section]
\newtheorem{propn}[thm]{Proposition}

\def\be{\begin{equation}}
\def\ee{\end{equation}}
\def\bea{\begin{eqnarray}}
\def\eea{\end{eqnarray}}

\def\dblone{\hbox{$1\hskip -1.2pt\vrule depth 0pt height 1.6ex width 0.7pt \vrule depth 0pt height 0.3pt width 0.12em$}}
\def\d'{``}

\begin{document}

\begin{center}
\Large{\bf{A q-difference Baxter's operator for the Ablowitz-Ladik chain \footnote{This is a shorter version of the same article published in {\it J. Phys. A: Math. Theor.} {\bf 48}, 2015, 125205. doi:10.1088/1751-8113/48/12/125205} }}
\end{center}

\begin{center}
{ {  Federico Zullo}}

{ Dipartimento di Matematica e Fisica, Universit\'a di Roma Tre\\ \&\\ INFN, Sezione di Roma Tre\\ Via della Vasca Navale 84, 00146, Roma, Italy
\\~~e-mail: zullo@fis.uniroma3.it}

\end{center}

\medskip
\medskip

\begin{abstract}
\noindent We construct the Baxter's operator and the corresponding Baxter's equation for a quantum version of the Ablowitz Ladik model. The result is achieved by looking at the quantum analogue of the classical B\"acklund transformations. For comparison we find the same result by using the well-known Bethe ansatz technique. General results about integrable models governed by the same $r$-matrix algebra will be given. The Baxter's equation comes out to be a q-difference equation involving both the trace and the quantum determinant of the monodromy matrix. The spectrality property of the classical B\"acklund transformations gives a trace formula representing the classical analogue of the Baxter's equation. An explicit q-integral representation of the Baxter's operator is discussed.

\end{abstract}

\bigskip\bigskip

\noindent


\section{Introduction.} \label{intr}
The Ablowitz-Ladik model \cite{AL0},\cite{AL1} is the integrable differential difference version of the nonlinear Schr\"odinger equation. The dynamical variables lie on a (periodic or infinite) lattice: we shall denote by a subscript \d'$k$'' the points of the lattice and by $(r_k,q_k)$ the dynamical variables. The classical equations of motion are given by
\begin{equation*}\begin{aligned}
&\dot{q}_k=q_{k+1}+q_{k-1}-2q_k-q_kr_k(q_{k+1}+q_{k-1}), \\
&\dot{r}_k=-r_{k+1}-r_{k-1}+2r_k+q_kr_k(r_{k+1}+r_{k-1}).
\end{aligned}
\end{equation*}  
This model possesses a Lax matrix representation. 
The Lax matrix is given by the product \cite{BIK}, \cite{Kulish}, \cite{FZAL}
\begin{equation}\label{Lm}
L(\lm)=\stackrel{\curvearrowleft}{\prod_{k=1}^N}L_k(\lm), \qquad \textrm{with} \qquad L_{k}(\lm) =  \left( \begin{array}{cc} \lm & q_k\\ r_k& \lambda^{-1}\end{array} 
\right).
\end{equation}
where 
$\lambda$ is the spectral parameter. The determinant of $L(\lambda)$, given by $\prod_{k=1}^{N}(1-q_{k}r_{k})$, is a conserved quantity. Other $N-1$ conserved quantities appear in the Laurent expansion of the trace of $L(\lambda)$
\begin{equation}\label{Trcons}
\textrm{Tr}(L(\lambda))=\sum_{i=0}^{N}H_{i}\lambda^{N-2i},\quad H_{0}=H_{N}=1.
\end{equation}

The involutivity of these conserved quantities follows from the Poisson structure underlying the model: it can be defined by a $r$-matrix relation satisfied by the Lax matrix
\be
\{L(\lambda)\otimes L(\nu)\}=[r,\ L(\lambda)\otimes L(\nu)],
\ee
where the $r$ matrix is defined by \cite{S}
$$
r(\lm,\nu)\doteq\left( \begin{array}{cccc} \frac{1}{2}\frac{\nu^2+\lm^2}{\nu^2-\lm^2} & 0&0&0\\ 0&-\frac{1}{2}&\frac{\lm \nu}{\nu^2-\lm^2}&0 \\ 0&\frac{\lm \nu}{\nu^2-\lm^2}&\frac{1}{2}&0\\0&0&0& \frac{1}{2}\frac{\nu^2+\lm^2}{\nu^2-\lm^2}  \end{array} \right).
$$

The above relations are equivalent to the following Poisson
brackets for the dynamical variables of the model
\be\label{PB}
\{q_k,r_j\}=(1-q_kr_k)\delta_{kj}, \qquad \{q_k,q_j\}=\{r_k,r_j\}=0.
\ee

The exact-discretization of this model through the technique of B\"acklund transformations 
has been considered recently in \cite{FZAL}, where it was also shown how it is possible
to get explicit maps preserving exactly the continuous trajectories. Those results were
obtained using the general scheme described in \cite{FZBH}, that, in turns, was stimulated
by earlier results on the application of the theory of B\"acklund transformations to finite
dimensional systems \cite{KS},\cite{RZK},\cite{RZTG},\cite{FZGC}.\\

In this work we consider the quantum version of the Ablowitz-Ladik model and the corresponding quantum version of its B\"acklund transformations. The quantum version of the Ablowitz-Ladik model, with the Lax and $r$-matrix structures, were introduced by Kulish \cite{Kulish}. Here we show how the quantum analogue of the classical B\"acklund transformations lead to the Baxter's equations for this model. Similar results have been obtained for the discrete self trapping model \cite{KSS} and for the Toda lattice \cite{PG}, \cite{S}.
The quantum Ablowitz Ladik model corresponds to the q-boson model \cite{BBP}, \cite{BIK}: the algebra (\ref{com}) defined by the $r$-matrix structure (\ref{quantumr})-(\ref{CR}) can be put in correspondence with the $q$-boson algebra \cite{BBP}, \cite{BBP1}, \cite{BIK}, \cite{Korff}, (also called $q$-oscillator algebra \cite{KSbook}). The operators $r_k$ and $q_k$ can be interpreted as boson creation/annihilation operators. The eigenvalue problem for the q-boson model with periodic boundary conditions is also equivalent to the eigenvalue problem of an integrable discretization \cite{VD1} of the Lieb-Liniger model \cite{LL} (see \cite{Korff}). For more details about the Bethe wave functions, the completeness of the Bethe ansatz, the correlation functions of the model, the properties of the model in the thermodynamic limit and results about the infinite chain we refer the reader to the papers \cite{BBP}, \cite{BIK}, \cite{BBP1}, \cite{Korff}, \cite{VD1}, \cite{VD} and references therein. In this paper, as said, a particular emphasis will be put on the application of the B\"acklund transformations theory to the quantum model. \\

In general, for a given integrable system, the quantum B\"acklund transformations can be represented by an unitary operator $\mathcal{Q}_{\mu}$, where $\mu$ is a (set of) parameter(s) of the transformations, realized as an integral operator on the space of
eigenfunctions \cite{KS}, \cite{PG}. If we denote by $(P,Q)$ a set of canonically conjugated variables
and by $(\tl{P},\tl{Q})$ the new variables transformed according to the B\"acklund maps, we have
\be 
\mathcal{Q}_{\mu}: \psi(Q)\to \int f(\tl{Q},Q)\psi(Q)dQ
\ee
The similarity transformations induced by $\mathcal{Q}_{\mu}$ are the equivalent of the classical canonical
transformations and the kernel $f(\tl{Q},Q)$ is given, in the semiclassical approximation,
by
\be
f(\tl{Q},Q)\sim \exp(-\frac{i}{\hbar}F(\tl{Q},Q)), \quad \hbar \to 0,
\ee
where $F(\tl{Q},Q)$ is the generating function of the classical B\"acklund transformations, that indeed are canonical transformations of the phase space. The $\mathcal{Q}$ operator is called the Baxter's operator. The explicit construction of a Baxter's operator for the model under consideration has been given recently by Korff \cite{Korff}: the reader can find in the corresponding paper (see Proposition 3.11) the explicit expression of the Baxter's operator in terms of polynomials in the generators of the q-boson algebra. Here, in the spirit of the above description, we obtain an integral formula for the Baxter's operator.\\ 

For integrable models possessing commutation relations among the dynamical variables defined by a given class of r-matrices, the
Baxter's operator solves a linear finite-difference spectral problem. More precisely, if
$r=\lm+\eta\mathcal{P}$, where $\mathcal{P}$ is the permutation operator in the tensor product $\mathbb{C}^2\otimes\mathbb{C}^2$
(i.e. $\mathcal{P} x\otimes y=y\otimes x \forall x,y \in \mathbb{C}^2$), then we have
\be\label{Bax}
\mathcal{Q}_{\mu}\textrm{Tr}(L(\mu))=\Delta_+\mathcal{Q}_{\mu+\eta}+\Delta_-\mathcal{Q}_{\mu-\eta}
\ee
where $\Delta_{\pm}$ are scalar functions \cite{B},\cite{KS},\cite{PG}. Equation (\ref{Bax}) is considered as a fundamental attribute of the
Baxter's operator and the starting point for further investigations. Here we emphasize also the significant relationships between this operator and the B\"acklund
transformations theory. Indeed we will see that in our case, the form of equation
(\ref{Bax}) is changed in a q-difference equation, whether the relationship with the theory of B\"acklund transformations remains unchanged.

\section{The Bethe ansatz and the Baxter's equation.}\label{Bethesec}
In this section we shall construct the Baxter's equation by diagonalizing the operator
Tr$(L(\mu))$. We will use the standard algebraic Bethe ansatz technique (see e.g. \cite{Faddeev}).\\

The quantum version of the Ablowitz-Ladik model is defined by the same Lax matrix as above (\ref{Lm}), where now the elements of the matrices are operator. The commutation relations are defined by the following quantum r-matrix (see also \cite{Kulish})
\begin{equation}\label{quantumr}
R(\lm /\nu)\doteq \left( \begin{array}{cccc} 1+\eta c & 0&0&0\\ 0&1+\eta&\eta b&0 \\ 0&\eta b&1&0\\0&0&0& 1+\eta c  \end{array} \right),\qquad \eta\doteq \i \hbar 
\end{equation}
where, for simplicity of notation, we set $c=\frac{\lm^2}{\lm^2-\nu^2}$ and $b=\frac{\lm\nu}{\lm^2-\nu^2}$. This quantum r-matrix is related to the classical one by the relation $R=\left(1+\eta/2\right)\dblone-\eta r$. The $r$-matrix (\ref{quantumr}) solves the Yang-Baxter equation
\begin{equation}\label{YB}
R_{ic,ja}(\lm /\nu)R_{cm,kb}(\lm)R_{an,br}(\nu)=R_{ja,kb}(\nu)R_{ic,br}(\lm)R_{cm,an}(\lm /\nu)
\end{equation}
In equation (\ref{YB}) there is a sum over repeated indices and we use the usual convention for the Kronecker product of two matrices to label the elements $R_{\alpha \beta, \gamma\delta}$ (i.e. if $T=A\otimes B$ then $T_{\alpha\beta, \gamma\delta}=A_{\alpha\beta}B_{\gamma\delta}$).

The Poisson brackets (quantum commutators) among phase space variables are defined
by the relation
\be\label{CR}
R(\lm /\nu)\stackrel{1}{L}(\lm)\stackrel{2}{L}(\nu)=\stackrel{2}{L}(\nu)\stackrel{1}{L}(\lm)R(\lm /\nu)
\ee
where $\stackrel{1}{L}$ and $\stackrel{2}{L}$ mean the tensor products $\stackrel{1}{L}=L\otimes\dblone$ and $\stackrel{2}{L}=\dblone\otimes L$.  The commutation relations (\ref{CR}) are equivalent to
\begin{equation}\label{com}
[q_j,r_k]=\eta (1-q_jr_j)\delta_{j,k},
\end{equation}
that are the quantum analogue of the Poisson brackets (\ref{PB}). The relations (\ref{CR}) define
the commutators among the matrix elements of the monodromy matrix $L(\lm)$. \\
By using recursively the commutation relations (\ref{CR}), if we set
\begin{equation}\label{mon}
L(\lm)=\left( \begin{array}{cc} A(\lm) & B(\lm)\\ C(\lm)& D(\lm)\end{array} 
\right),
\end{equation}
then it is possible to state the following proposition:
\begin{propn}
Defining a pseudo-vacuum state $|0\rangle$ by $B(\lm)|0\rangle=0$, the eigenfunctions of Tr$(L(\nu))$ are defined by
\begin{equation}\label{eigenf}
\phi(\{\lm\})=\prod_{k=1}^m C(\lm_k)|0\rangle , 
\end{equation}
where the constants $\lm_k$ are defined by the expressions
$$
\prod_{j\neq k}\left(\frac{\lm_j^2(1+\eta)-\lm_k^2}{\lm_j^2-(1+\eta)\lm_k^2}\right)=\lm_k^{2N}, \quad k=1\dots m,
$$ 
and the eigenvalue equations are given by
\begin{equation}\label{eig}
\textrm{Tr}(L(\nu))\phi=\frac{\nu^N}{(1+\eta)^m}\prod_{j=1}^m \left(1-\eta\frac{\nu^2}{\lm_j^2-\nu^2}\right)\phi+\frac{\nu^{-N}}{(1+\eta)^m}\prod_{j=1}^m \left(1+\eta\frac{\lm_j^2}{\lm_j^2-\nu^2}\right)\phi .
\end{equation}
\end{propn}
If we set
$$
\psi(\nu,\{\lm\})\doteq \prod_{j=1}^m(\nu^2-\lm_j^2), \qquad t(\nu)=\textrm{Tr}\left(L(\nu)\right),
$$
then equation (\ref{eig}) is equivalent to
\begin{equation}\label{trax1}
t(\nu)\psi(\nu,\{\lm\})=\frac{\nu^N}{(1+\eta)^m}\psi(\nu\sqrt{\eta+1},\{\lm\})+\frac{1}{\nu^N}\psi(\frac{\nu}{\sqrt{1+\eta}},\{\lm\})
\end{equation}
To better understand the structure of this equation we introduce the quantum determinant of the monodromy matrix (\ref{mon}) (see \cite{FZ} or \cite{Kulish}). For clarity of notation let us set
\begin{equation}\label{alpha}
\alpha\doteq\frac{1}{1+\eta}=\frac{1}{1+\textrm{i}\hbar},
\end{equation}
then the quantum determinant $\Delta$ of (\ref{mon}) is given by
\begin{equation}\begin{split}\label{qdet}
\left(\sqrt{\alpha}\right)^{N-1}\Delta & \doteq \frac{A(\lm)D(\lm\sqrt{\alpha})}{\sqrt{\alpha}}-\frac{B(\lm)C(\lm\sqrt{\alpha})}{\alpha}= \frac{D(\lm)A(\lm\sqrt{\alpha})}{\sqrt{\alpha}} - C(\lm)B(\lm\sqrt{\alpha})=\\
& =\frac{A(\lm\sqrt{\alpha})D(\lm)}{\sqrt{\alpha}}-C(\lm\sqrt{\alpha})B(\lm)=\frac{D(\lm\sqrt{\alpha})A(\lm)}{\sqrt{\alpha}}-\frac{B(\lm\sqrt{\alpha})C(\lm)}{\alpha}.
\end{split}\end{equation}
These four expressions are equal by virtue of the commutation relations (\ref{CR}). 
The following proposition can be easily proved \cite{FZ}
\begin{propn}
The operators $\Delta(\lm)$ and $\textmd{Tr}(L(\mu))$ commute. The eigenvalues of the quantum determinant $\Delta$ (\ref{qdet}) on the eigenfunctions (\ref{eigenf}) are given by the relation
$$
\Delta(\lm)\prod_{j=1}^m C(\lm_j)|0\rangle = \alpha^m\frac{d(\lm)a(\lm\sqrt{\alpha})}{\left(\sqrt{\alpha}\right)^{N}}\prod_{j=1}^m C(\lm_j)|0\rangle.
$$
\end{propn}

The previous formulae and proposition are independent of the particular structure of the monodromy matrix.
If we specialize to our monodromy matrix, defined by (\ref{Lm}), $\Delta$ 
can be explicitly written in terms of the dynamical variables $\{q_k, r_k\}_{k=1}^{N}$ as \cite{FZ}
\begin{equation}\label{qdet1}
\Delta=\prod_{k=1}^N (1-r_kq_k)
\end{equation}
The corresponding eigenvalues are independent of $\lm$ and are given by $\alpha^m$.\\
The meaning of equation (\ref{trax1}) is now clearer: the trace decomposes in the sum of two terms that classically correspond to the two eigenvalues of the monodromy matrix. If we denote by $\delta$ the eigenvalue of the quantum determinant $\Delta$, we have 
\begin{equation}\label{trax}
t(\nu)\psi(\nu,\{\lm\})=\delta\nu^N\psi(\frac{\nu}{\sqrt{\alpha}},\{\lm\})+\frac{1}{\nu^N}\psi(\nu\sqrt{\alpha},\{\lm\})
\end{equation}
By choosing a different normalization for the function $\psi(\nu)$, that is $\psi(\nu)=\hat{\psi}(\nu)\nu^{2m}$, one gets
\begin{equation}\label{traxm}
t(\nu)\hat{\psi}(\nu,\{\lm\})=\nu^N\hat{\psi}(\frac{\nu}{\sqrt{\alpha}},\{\lm\})+\frac{\delta}{\nu^N}\hat{\psi}(\nu\sqrt{\alpha},\{\lm\})
\end{equation}
that is the factor $\delta$ moved to one addend to the other.

In general we expect the Baxter's equation defined by monodromy matrices satisfying the commutation relation (\ref{CR}) to be
\begin{equation}\label{gentrax}
t(\nu)\psi(\nu,\{\lm\})=\Lambda_+\psi(\frac{\nu}{\sqrt{\alpha}},\{\lm\})+\Lambda_{-}\psi(\nu\sqrt{\alpha},\{\lm\})
\end{equation}
where $\Lambda_+$ and $\Lambda_{-}$ are two scalar factors whose product gives $\delta$, the eigenvalue of the corresponding quantum determinant. A similar functional equation has been obtained by Korff for the $\mathcal{Q}$ operator  (see \cite{Korff}, eq. 3.50).

\section{Baxter's equation and B\"acklund transformations.}\label{Gen}
In this section we find an expression for the Baxter's $\mathcal{Q}$ operator investigating the relationships with a set of B\"acklund transformations for the model. We will divide the section in three parts: the first one will describe a set of classical B\"acklund transformations for the model, mainly emphasizing the aspects having a clear correspondence with the quantum case, that will be described in the second part of the section. In the last subsection we represent the Baxter's operator as a q-integral operator, giving its main properties (i.e. the Baxter's equation and the commutativity with the conserved quantities of the model). For all the details we refer the reader to \cite{FZ}. 

\subsection{Classical B\"acklund transformations} \label{Classical}
Classical B\"acklund transformations for the Ablowitz-Ladik model have been considered in different works (see e.g. \cite{S}, \cite{FZAL}). It is possible to compose elementary parametric transformations to get more complex multi-parametric maps. Here, for what concerns our purposes, it is sufficient to take just an elementary transformation. We shall adapt the results of \cite{FZAL}. Our set of B\"acklund transformations can be found from the matrix equation
\begin{equation}\label{LtDDL}
\tl{L}_k\mathcal{D}_k-\mathcal{D}_{k+1}L_k=0
\end{equation}
where the dressing matrix $\mathcal{D}_k$ is given by (see \cite{FZ} for more details)
\begin{equation}\label{claD}
\mathcal{D}_{k}(\lm) =  \left( \begin{array}{cc} \lm^2-\mu^2(1-b_kc_{k}) & \lm b_k\\ \lm c_{k}& 1\end{array} 
\right),
\end{equation} 
with $b_k=q_k$ and $c_k=\tl{r}_{k-1}$. The transformations can be written as
\begin{equation}\label{BTs}
\begin{aligned}
&1-q_k r_k=\frac{(\tl{r}_{k-1}-r_{k})(\tl{r}_k\mu^2+r_k)}{\mu^2\tl{r}_{k}\tl{r}_{k-1}},\\
&1-\tl{q}_k\tl{r}_k=\frac{(\tl{r}_k-r_{k+1})(\tl{r}_k\mu^2+r_k)}{\mu^2\tl{r}_{k+1}\tl{r}_{k-1}},
\end{aligned}\qquad k=1\ldots N
\end{equation}
defining $\tl{q}_k$ and $q_k$ in terms of the independent variables. We are assuming periodic boundary conditions, i.e. $q_{k+N}=q_k$ and $r_{k+N}=r_{k}$. These expressions do define B\"acklund transformations for the Ablowitz-Ladik model because
\begin{enumerate}
\item the conserved quantities are invariant under the action of the map (\ref{BTs}), 
\item the transformations are canonical.
\end{enumerate}

For the point 2 it is possible to write down explicitly the generating function of the transformations. Indeed, since $-\frac{\ln(1-q_nr_n)}{r_n}dr_n$ is a canonical one-form, the generating function must satisfy
\begin{equation}\label{gfun}
dF(r,\tl{r})=\sum_{k}\frac{\ln(1-\tl{q}_k\tl{r}_k)}{\tl{r}_k}d\tl{r}_k-\frac{\ln(1-q_kr_k)}{r_k}dr_k .
\end{equation} 
The function $F(r,\tl{r})$ is given by
$$
F=\sum_k \int_{r_{k+1}+1}^{\tl{r}_k}\frac{\ln(z-r_{k+1})}{z}dz+\int_{1/\mu^2}^{\tl{r}_k}\frac{\ln(\mu^2z+r_{k})}{z}dz-\ln(\tl{r}_k)\ln(\mu^2\tl{r}_{k-1})-2\ln(\mu)^2.
$$

Now we shall show how it is possible to obtain the classical version of the Baxter's equation (\ref{gentrax}) by looking at the action of the dressing matrix on the matrices $L_k(\lm)$. The result will be trivial, but it will help to understand what we get in the quantum case. The idea is a mix of the observations due to Pasquier \& Gaudin \cite{PG} and Kuznetsov \& Sklyanin \cite{KS}. The Baxter's equation (\ref{gentrax}) involves the trace of the monodromy matrix so we would like to obtain an expression for this trace. The observation in \cite{PG} is that Tr$L(\mu)$ doesn't change if we perform a sort of similarity transformation on the matrices $L_k$ like $\hat{L}_k=M_{k+1}^{-1}L_kM_{k}$. To choose the matrix $M_k$ one has to look at another matrix, i.e. the dressing matrix defining the B\"acklund transformations \cite{FZ}. It depends on the spectral parameter $\lm$ but when $\lm$ is equal to the parameter $\mu$ of the B\"acklund transformations then it becomes singular, its kernel being given by 
$$
|w_k\rangle=\left( \begin{array}{c} 1 \\ -\mu\tl{r}_{k-1}\end{array} 
\right).
$$
From the isospectral equation satisfied by the Lax matrix $L_k$ and the dressing matrix it follows that
\begin{equation}\label{eigs}
L_k(\mu)|w_k\rangle=\gamma_k |w_{k+1}\rangle
\end{equation}
for some function $\gamma_k$. We notice that $|w_1\rangle$ is an eigenvector of the monodromy matrix $L(\mu)$ and $\gamma=\prod_{k=1}^N \gamma_k$ is the corresponding eigenvalue. The other eigenvalue is given by $\frac{\det(L(\mu))}{\gamma}=\prod_{k=1}^N\frac{1-q_kr_k}{\gamma_k}$. We build a matrix $M_k$ with the column $|w_k\rangle$ and another arbitrary column $|x_k\rangle$:
$
M_k=\left(|x_k\rangle,|w_k\rangle\right).
$
With the help of relation (\ref{eigs}) one can show that $\hat{L}_k$ is lower triangular
\begin{equation}\label{trian}
\hat{L}_k=M_{k+1}^{-1}L_kM_{k}=\left( \begin{array}{cc} \frac{\det(M_k)}{\det(M_{k+1})}\frac{1-q_kr_k}{\gamma_k} & 0\\ *& \gamma_k\end{array} 
\right)
\end{equation}
immediately implying
\begin{equation}\label{cb}
\textrm{Tr}\hat{L}(\mu)=\textrm{Tr}L(\mu)=\prod_{k=1}^N\frac{1-q_kr_k}{\gamma_k}+\prod_{k=1}^N\gamma_k=\frac{\det(L(\mu))}{\gamma}+\gamma.
\end{equation}
Although the result is trivial (the trace of the monodromy matrix is the sum of its eigenvalues), there is a strong correspondence with the quantum case. Indeed (\ref{cb}) is the semi-classical limit of the Baxter's equation, as we will show in the next subsection. Now we will just make some further considerations, postponing more comments to the end of subsection \ref{three}. 

The parameter $\mu$ can be seen as an evolution parameter for the B\"acklund transformations. Then the maps are the integral curves of a non autonomous Hamiltonian system of equations, the flow being generated by the variable conjugated to $\mu$ expressed in the variables $(r,q)$ \cite{FZBH}-\cite{FZAL}, that is by $\Phi=\left.\frac{\partial F}{\partial \mu}\right|_{\tl{r}=\tl{r}(r,q)}$. Explicitly we have
\begin{equation}\label{Ph}
\Phi=\left.\frac{\partial F}{\partial \mu}\right|_{\tl{r}=\tl{r}(r,q)}=\left.\frac{2}{\mu}\sum_k \ln\left(\frac{\mu^2\tl{r}_k+r_k}{\mu^2\tl{r}_k}\right)\right|_{\tl{r}=\tl{r}(r,q)}.
\end{equation}
We notice that $\Phi$ can be written as
\begin{equation}\label{Phi}
\Phi=\left.\frac{2}{\mu}\sum_k \ln\left(\frac{\mu^2\tl{r}_k+r_k}{\mu^2\tl{r}_k}\right)\right|_{\tl{r}=\tl{r}(r,q)}=\frac{2}{\mu}\ln \prod \frac{(1-q_kr_k)}{\mu\gm_k}=\frac{2}{\mu}\ln\left( \frac{
 \det(L(\mu))}{\mu^N\gm}\right)
\end{equation}
Comparing with equation (\ref{cb}), we see that it can be represented as
\begin{equation}\label{cltr}
\textrm{Tr}L(\mu)=\mu^N e^{\frac{\mu}{2}\Phi}+\frac{\det(L(\mu))}{\mu^N}e^{-\frac{\mu}{2}\Phi}.
\end{equation}
We shall return on this equation at the end of the next subsection.

\subsection{Quantum case}\label{Quantum}
The quantum model is described by the monodromy matrix (\ref{Lm}) and the commutation relations (\ref{CR}). The action of $q_k$ on a function $f(\{r_j\})$ is proportional to the Jackson derivative (or $q$-derivative, but to avoid confusion with the dynamical variables we shall use the symbol $\alpha$ instead of $q$) in the direction of $q_k$ (see also \cite{KSbook}, Chapter 5)
\begin{equation}\label{jack}
q_k f(\{r_j\})=(1-\alpha)D_{\alpha,k}f(\{r_j\}), \quad D_{\alpha,k}f(\{r_j\})\doteq\frac{f(r_1,\dots,\alpha r_k,\dots,r_N)-f(r_1,\dots,r_k,\dots,r_N)}{\alpha r_k-r_k}.
\end{equation}
where $\alpha$ is defined by (\ref{alpha}).\\

Again we shall follow \cite{PG} and \cite{KS} (see also \cite{B}). We are looking for an operator $\mathcal{Q}_\mu$ satisfying (see eq. (\ref{gentrax}))
\begin{equation}\label{Baxter}
\textrm{Tr}(L(\mu))\mathcal{Q}_\mu=\mu^N\mathcal{Q}_{\frac{\mu}{\sqrt{\alpha}}}+\frac{\Delta}{\mu^N}\mathcal{Q}_{\mu\sqrt{\alpha}}, \quad \alpha\doteq \frac{1}{1+\eta}=\frac{1}{1+\i \hbar}
\end{equation}
commuting with the trace of $L(\lm)$, $[\mathcal{Q}_\mu,\textrm{Tr}L(\lm)]=0$. \\
Let us consider the columns of equation (\ref{Baxter}), say $\rho$. If we take $\rho$ in the form of a product, i.e. $\rho=\prod_j \rho_j(r_j)$, then, because $L(\mu)$ is itself in the form of a product (see eq. (\ref{Lm})),  the action of $L(\mu)$ on $\rho$ decomposes:
\begin{equation}\label{prodmon}
\textrm{Tr}L(\mu)\rho=\textrm{Tr}\left((L_N\rho_N) ... (L_1\rho_1)\right)
\end{equation} 
Exactly as in the classical case, Tr$L(\mu)$ doesn't change if we perform the transformation  $\hat{L}_k=M_{k+1}^{-1}L_kM_{k}$. So we take the following matrix $M_k$:
$$
M_k=\left( \begin{array}{cc} 0 & 1\\ -1& -\mu \tl{r}_{k-1}\end{array} 
\right)
$$
that is of the same form as the one in the classical case, but now we specified the vector $|x_k\rangle$ to be $(0,1)^{T}$. We fixed this vector for the sake of simplicity since now the determinant of the matrix $M_k$ is equal to 1. We get
\begin{equation}\label{Lt}
\hat{L}_k=\left( \begin{array}{cc} \mu \tl{r}_kq_k+\frac{1}{\mu} & -\mu^2\tl{r}_k(1-q_k\tl{r}_{k-1})-(r_k-\tl{r}_{k-1})\\ -q_k& \mu(1-q_k\tl{r}_{k-1}) \end{array} 
\right)
\end{equation}
The functions $\rho_k$ are then defined by requiring (see eq. (\ref{trian}))
\begin{equation}\label{cond}
\mu^2\tl{r}_k(1-q_k\tl{r}_{k-1})\rho_k=(\tl{r}_{k-1}-r_k)\rho_k
\end{equation}
Recalling the action of the operator $q_k$ on the functions of $r_k$ (\ref{jack}), equation (\ref{cond}) can be written as
$$
\rho_k(\mu,r_k)=\frac{\mu^2\tl{r}_k\tl{r}_{k-1}}{(\mu^2\tl{r}_k+r_k)(\tl{r}_{k-1}-r_k)}\rho_k(\mu,\alpha r_k)
$$
that is solved by
\begin{equation}\label{rhok}
\rho_k(\mu,r_k)=G_k\prod_{p=0} \frac{1}{(1+\alpha^p \frac{r_k}{\mu^2\tl{r}_k})(1-\alpha^p \frac{r_k}{\tl{r}_{k-1}})}=\frac{G_k}{(\frac{r_k}{\tl{r}_{k-1}};\alpha)_{\infty}(-\frac{r_k}{\mu^2\tl{r}_{k}};\alpha)_{\infty}} 
\end{equation}
where $G_k$ is independent of $r_k$ and we introduced the usual notation for the $q$-Pochhammer symbol \cite{BHS}
\begin{equation}\label{Pochh}
(x;\alpha)_{\infty}=\prod_{p=0}^{\infty}(1-x\alpha^p).
\end{equation}
Notice that $\frac{1}{(x;\alpha)_{\infty}}$ is a q-analog of the exponential function $e^x$ since $\frac{1}{(x(1-\alpha);\alpha)_{\infty}}\to e^x$ in the limit $\alpha\to 1$.\\
Using (\ref{cond}), the matrix $\hat{L}(\mu)\rho_k$ can be written as
\begin{equation}\label{Lt1}
\hat{L}_k\rho_k=\left( \begin{array}{cc} \frac{\mu^2\tl{r}_k+r_k}{\mu\tl{r}_{k-1}} & 0\\ -q_k& \frac{\tl{r}_{k-1}-r_k}{\mu\tl{r}_k} \end{array} 
\right)\rho_k
\end{equation}
or, using the explicit form of the functions $\rho_k$ (\ref{rhok})
\begin{equation*}
\hat{L}_k\rho_k(\mu,r_k)=\left( \begin{array}{cc} \frac{\mu\tl{r}_k}{\tl{r}_{k-1}}\rho_k(\frac{\mu}{\sqrt{\alpha}},r_k) & 0\\ -q_k \rho_k(\mu,r_k)& \frac{\tl{r}_{k-1}}{\mu\tl{r}_k}\rho_k(\mu\sqrt{\alpha},\alpha \rho_k) \end{array} 
\right).
\end{equation*}
Then it follows from (\ref{prodmon}) that the trace of the monodromy matrix $L(\mu)$ on $\rho=\prod_k \rho_k$ is given by
\begin{equation}\label{Trint}
\textrm{Tr}(L(\mu))\rho(\mu,r)=\mu^N\rho(\frac{\mu}{\sqrt{\alpha}},r)+\frac{1}{\mu^N}\rho(\mu\sqrt{\alpha},\alpha r).
\end{equation}
The action of $\Delta=\prod_k (1-r_kq_k)$ on $\rho$ is given by
$$
\prod_k (1-r_kq_k)\rho(r)=\rho(\alpha r)
$$
so that equation (\ref{Trint}) can be also written as
\begin{equation}\label{qtr}
\textrm{Tr}(L(\mu))\rho(\mu,r)=\mu^N\rho(\frac{\mu}{\sqrt{\alpha}},r)+\frac{\Delta}{\mu^N}\rho(\mu\sqrt{\alpha},r).
\end{equation}

From the B\"acklund transformations point of view there are clear analogies between the classical and the quantum cases: the Baxter's equation
\begin{equation}\label{qtr1}
\textrm{Tr}(L(\mu))\mathcal{Q}_\mu=\mu^N\mathcal{Q}_{\frac{\mu}{\sqrt{\alpha}}}+\frac{\Delta}{\mu^N}\mathcal{Q}_{\mu\sqrt{\alpha}}
\end{equation}
is the quantum analogue of the classical equation (\ref{cltr}) for the trace of the monodromy matrix. Indeed in (\ref{cltr}) $\Phi$ is the canonical conjugated variable with respect to $\mu$. If we substitute directly for $\Phi$ the corresponding quantum variable, i.e. $\Phi=\eta\frac{\partial}{\partial\mu}$, then equation (\ref{cltr}) gives
\begin{equation}
\begin{split}
\textrm{Tr}L(\mu)\rho(\mu,r)&=\mu^N e^{\frac{\mu}{2}\eta\frac{\partial}{\partial\mu}}\rho(\mu,r)+\frac{\Delta}{\mu^N}e^{-\frac{\mu}{2}\eta\frac{\partial}{\partial\mu}}\rho(\mu,r)=\\
&=\mu^N\rho(\mu(1+\frac{\eta}{2}))+\frac{\Delta}{\mu^N}\rho(\mu(1-\frac{\eta}{2})).
\end{split}
\end{equation}
that agrees at first order in $\eta$ with (\ref{qtr}) since $\alpha=\frac{1}{1+\eta}$. The fact that the semi-classical limit of the quantum B\"acklund transformations is linked with the generating function of the corresponding classical transformations is well known \cite{PG}, \cite{Sk}. If we consider the equation (\ref{qtr1}) for its eigenvalues
\begin{equation}\label{eigns}
t(\mu)q(\mu)=\mu^N q\left({\frac{\mu}{\sqrt{\alpha}}}\right)+\frac{\delta}{\mu^N} q\left({\mu\sqrt{\alpha}}\right)
\end{equation}
and we seek for a solution $q(\mu)$ in the following form
$$
q(\mu)=e^{\frac{1}{\eta}\left(S_0+\eta S_1+\eta^2 S_2+\ldots\right)},
$$
we get for $S_0$ (the prime indicates derivative with respect to $\mu$)
\begin{equation}\label{ford}
t(\mu)=\mu^N e^{\frac{\mu}{2}S_0'}+\frac{\det(L(\mu))}{\mu^N}e^{-\frac{\mu}{2}S_0'}.
\end{equation}
Confronting with equation (\ref{cltr}) and (\ref{Ph}) we see that $S_0=F$, the generating function of the classical transformations.

Let us give a remark. Usually the Baxter's equation involves two operators, that is the trace of the monodromy matrix and the Baxter's operator. This because the quantum determinant of the monodromy matrix is a c-number or a Casimir of the system. In this case however it is a true operator, so the Baxter's equation involves three operators commuting each other and with themselves. However it is possible to show \cite{FZ} that this single equation is still enough to get the eigenvalues of both the trace of the monodromy matrix and of its quantum determinant.

\subsection{A q-integral formula for the Baxter's operator}\label{three}
For quantum models with commutation relations described by the $r$-matrix given by the permutation operator (see eq. (\ref{Bax})), it is possible to give an explicit integral formula for the Baxter's operator \cite{S}. The kernel of the integral is the object related to the B\"acklund transformations, since it gives, in the semi-classical limit, the generating function of the classical transformations \cite{KS}. Further, the Baxter's operator can be expressed as the trace of a monodromy matrix \cite{B}, \cite{BLZ}, \cite{P}. It is possible to combine these two approaches \cite{KSS}: the monodromy matrix is again the product of elementary operators $\mathcal{R}_{\mu}^k$ from the spaces $\mathbb{C}[\tl{r}_k,c_k]$ to $\mathbb{C}[r_k,c_{k+1}]$, where $\mathbb{C}[c_k]$ are the spaces of the so-called \d'auxiliary variables''. The monodromy operator $\mathcal{R}_{\mu}$ is then a map from $\mathbb{C}[\tl{r},c_1]$ to $\mathbb{C}[r,c_{1}]$. Since in our case the Baxter's equation is a q-difference equation, we expect that the Baxter's operator can be represented by a q-integral formula. So we introduce the inverse of the operator $q_k$ defined in (\ref{jack})   
$$
(q_k)^{-1} f(\{r_j\})\doteq \int d_{\alpha}r_k f(\{r_j\}) \doteq \sum_{n=0}\alpha^n r_k f(r_1,\dots,\alpha^n r_k,\dots,r_N).
$$
Notice that $(1-\alpha)(q_k)^{-1}$ is the usual Jackson integral, the inverse of the Jackson derivative $D_{\alpha,k}f(\{r_j\})$ defined in formula (\ref{jack}). From the definition of the definite Jackson integral (see e.g. \cite{KC}) we also have
$$
\int_{0}^b d_{\alpha}r_k f(\{r_j\}) = \sum_{n=0}\alpha^n b f(r_1,\dots,\alpha^n b,\dots,r_N)
$$
and $\int_{a}^b d_{\alpha}r_k f(\{r_j\})\doteq \int_{0}^b d_{\alpha}r_k f(\{r_j\})-\int_{0}^a d_{\alpha}r_k f(\{r_j\})$. The calculus rules for $q_k$ and $(q_k)^{-1}$ then follow directly from the well-known q-calculus rules (see e.g. \cite{KC}). In particular we shall need of the q-Leibniz rule, given by
\begin{equation}\label{Leib}
q_k \Big(f(\{r_j\}) g(\{r_j\})\Big)=f(\{r_j\})q_k g(\{r_j\})+g(r_1,\dots,\alpha r_k,\dots,r_N)q_k f(\{r_j\}),
\end{equation}
and of the integration by parts
\begin{equation}\label{parts}
\int_{a}^b d_{\alpha}r_k f(\{r_j\})\left(q_k g(\{r_j\})\right)=f_k(b)g_k(b)-f_k(a)g_k(a)-\int_{a}^b d_{\alpha}r_k g(r_1,\dots,\alpha r_k,\dots,r_N)\left(q_k f(\{r_j\})\right)
\end{equation}
where we used the shorthand notation $f_k(b)=(r_1,\dots, b,\dots,r_N)$, that is $f_k(b)$ is the function $f(\{r_j\})$ with the variable $r_k$ replaced by $b$ and the same for $g_k(a)$. 

We introduce the following q-integral representation for the operators $R_{\mu}^k$
$$
\mathcal{R}_{\mu}^k: \psi(c_k,\tl{r}_k) \to \int d_{\alpha} c_k\int d_{\alpha}\tl{r}_k P_k(\alpha c_k,\alpha \tl{r}_k | c_{k+1}, r_k) \psi(c_k, \tl{r}_k)
$$
where the limits of integration are determined by the range of variations of our variables: for definiteness we can think to integrals between $-1$ and $1$. Also we assume that the function $\psi$ vanishes on these boundaries. 

The $\mathcal{Q}$ operator is defined to be
\begin{equation}\label{Bop}
\mathcal{Q}_{\mu}: \psi(\tl{r}) \to \int d_{\alpha}\tl{r}_N\ldots \int d_{\alpha}\tl{r}_1 \hat{Q}_{\mu}(\alpha \tl{r} | r) \psi(\tl{r})
\end{equation}
where the kernel $\hat{Q}_{\mu}(\alpha \tl{r} | r)$ is given by
\begin{equation}\label{hatQ}
\hat{Q}_{\mu}(\alpha \tl{r} | r)=\int d_{\alpha}c_N \ldots \int d_{\alpha}c_1  \prod_{k=1}^N P_k(\alpha c_k,\alpha \tl{r}_k | c_{k+1}, r_k).
\end{equation}

The factor $\alpha$ multiplying the variables $c_k$ and $\tl{r}_k$ is just for convention: it will result useful when we need to apply the formula for integration by parts (\ref{parts}). The commutation between Tr$(L(\lm))$ and $\mathcal{Q}_{\mu}$ is ensured by the quantum analogue of the relation (\ref{LtDDL}), that is \cite{KSS}, \cite{FZ}
\begin{equation}\label{qLtDDL}
\mathcal{R}_{\mu}^k \tl{L}_k(\lm)\mathcal{D}_k(\lm,\mu)=\mathcal{D}_{k+1}(\lm,\mu) L_k(\lm)\mathcal{R}_{\mu}^k
\end{equation}
The equivalence (\ref{qLtDDL}) gives the following set of equations for $\mathcal{R}_{\mu}^k$
\begin{equation}\label{Req}\begin{split}
&\mathcal{R}_{\mu}^k \tl{r}_k=c_{k+1}\mathcal{R}_{\mu}^k,\quad \mathcal{R}_{\mu}^kb_k=q_k \mathcal{R}_{\mu}^k,\\
&\mathcal{R}_{\mu}^k \tl{q}_k=\left(b_{k+1}-\mu^2(1-b_{k+1}c_{k+1})q_k\right)\mathcal{R}_{\mu}^k, \quad \mathcal{R}_{\mu}^k\left(c_k-\mu^2\tl{r}_k(1-b_kc_k)\right)=r_k\mathcal{R}_{\mu}^k,\\
&\mathcal{R}_{\mu}^k\left(\tl{q}_kc_k-\mu^2(1-b_kc_k)\right)=\left(b_{k+1}r_k-\mu^2(1-b_{k+1}c_{k+1})\right)\mathcal{R}_{\mu}^k.
\end{split}\end{equation}
The first equation gives a contribution proportional to a q-delta function $\delta(\tl{r}_k-c_{k+1})$ to $P_k(\alpha c_k,\alpha \tl{r}_k | c_{k+1}, r_k)$, that is we can set $P_k(\alpha c_k,\alpha \tl{r}_k | c_{k+1}, r_k)=\delta_{\alpha}(\tl{r}_k-c_{k+1})F_k(\alpha c_k, c_{k+1}, r_k)$. Here the q-delta function is defined by the identity 
\begin{equation}\label{delta}
\int d_{\alpha}x \delta_{\alpha}(x-y)g(y)=g(x).
\end{equation}
We don't enter into the discussions about a formal definition of the q-delta function but we shall just use equation (\ref{delta}) as a working identity to find an explicit expression for the kernel $\hat{Q}_{\mu}$ (\ref{hatQ}). By using the q-Leibniz rule (\ref{Leib}) and the q-integration by parts (\ref{parts}) we get from (\ref{Req}) the following set of equations for $F_k(\alpha c_k,c_{k+1},r_k)$:
\begin{equation}\label{Feq}\begin{split}
& r_k F_k(\alpha c_k,c_{k+1},r_k)+\mu^2\alpha c_{k+1}F_k(\alpha c_k,\alpha c_{k+1},\alpha r_k)=(r_k+\mu^2\alpha c_{k+1})F_k(\alpha c_k,\alpha c_{k+1},r_k),\\
& (r_k-c_k)F_k(\alpha c_k,c_{k+1},r_k)=r_k F_k(c_k,c_{k+1},r_k)-c_kF_k(\alpha c_k,c_{k+1},\alpha r_k),\\
& c_k F_k(\alpha c_k,c_{k+1},r_k)=(r_k+\mu^2 \alpha c_{k+1})F_k(\alpha c_k,\alpha c_{k+1},r_k),\\
& (c_k-r_k)F_k(\alpha c_k,c_{k+1},r_k)=\mu^2 c_{k+1}F_k(c_k,c_{k+1},r_k)
\end{split}\end{equation}
It is possible to check that these equations are indeed compatible. The first and the third equation of the set together gives
$$
F_k(\alpha c_k,\alpha c_{k+1},r_k)=\frac{\mu^2\alpha c_k c_{k+1}}{(r_k+\mu^2\alpha c_{k+1})(c_k-r_k)}F_k(\alpha c_k,\alpha c_{k+1},\alpha r_k).
$$
Since this equation involves only differences in the variable $r_k$, we expect that it is compatible with equation (\ref{rhok}) for the function $\rho_k$ that we found in the previous section. Indeed, if we set
$$
F_k(\alpha c_k, c_{k+1},r_k)=A_{k}G_k(\alpha c_k, c_{k+1})\prod_{p=0} \frac{1}{(1+\alpha^p \frac{r_k}{\mu^2 c_{k+1}})(1-\alpha^p \frac{r_k}{c_k})},
$$
where $A_{k}$ are constants depending only on $\alpha$ and $\mu$, equations (\ref{Feq}) reduce to
\begin{equation}\label{GG}
G_k(\alpha c_k, c_{k+1})=\frac{\mu^2\alpha c_{k+1}}{c_k}G_k(\alpha c_k, \alpha c_{k+1}),\quad  G_k(\alpha c_k, c_{k+1})=\frac{\mu^2 c_{k+1}}{c_k}G_k( c_k, c_{k+1})
\end{equation}
These two equations imply that $G_k(c_k, c_{k+1})$ is an homogeneous function of its arguments of degree $-1$, that is $G_k( c_k, c_{k+1})=\alpha G_k(\alpha c_k, \alpha c_{k+1})$. We can use this homogeneity to reduce the two equations (\ref{GG}) to a single equation. Let us set
$$
G_k( c_k, c_{k+1})=\frac{1}{c_{k+1}}\left(\frac{c_k}{c_{k+1}}\right)^{\frac{2\ln(\mu)}{\ln(\alpha)}}\hat{G}_k(\frac{c_k}{c_{k+1}}).
$$
If we introduce $z_k=\frac{c_k}{c_{k+1}}$, equations (\ref{GG}) give
$$
\hat{G}_k(z_k)=z_k\hat{G}_k(\alpha z_k).
$$
Notice that in formula (\ref{hatQ}) only the product over $k$ is involved, so, recalling the periodic conditions $c_{N+j}=c_{j}$ we can set the product $\prod_k \hat{G}_k$ equal to a constant. Putting all together, formula (\ref{hatQ}) becomes
\begin{equation}
\hat{Q}_{\mu}(\alpha \tl{r} | r)= A\prod_{k=1}^N \frac{1}{\tl{r}_k(\frac{r_k}{\tl{r}_{k-1}};\alpha)_{\infty}(-\frac{r_k}{\mu^2\tl{r}_{k}};\alpha)_{\infty}} 
\end{equation}
where the q-Pochhammer symbol $(x;\alpha)_{\infty}$ is defined by (\ref{Pochh}) and $A$ is a constant of normalization that can depend on $\alpha$ and $\mu$.

\section{Discussion}
In this paper we considered a quantum version of the Ablowitz-Ladik model. We built the Baxter's equation in two different ways: by using the algebraic Bethe ansatz technique (section \ref{Bethesec}) and with the help of the quantum analogue of the classical B\"acklund transformations (subsection \ref{Quantum} and \ref{three}). Our main aim was to underline the deep relationships between the B\"acklund transformations and the Baxter's operator. The Baxter's equation turns out to be a q-difference equation (see eq. \ref{qtr1}) whose semi-classical limit is linked with the classical trace formula for the monodromy matrix and with the spectrality property of the classical B\"acklund transformations (see eq. (\ref{ford})). The quantum determinant of the monodromy matrix is a conserved quantity but not a Casimir of the Poisson algebra defined by the commutation relations (\ref{CR}) and it plays an explicit role in the Baxter's equation. The construction leading to formulae (\ref{BTs}) for the classical B\"acklund transformations has a well defined and precise quantum counterpart, described in subsection \ref{Quantum}. In subsection \ref{three} we gave a q-integral formula for the Baxter's operator and proved the commutativity properties of $\mathcal{Q}_\mu$ with the other conserved quantities of the model encoded into the trace and quantum determinant of the monodromy matrix. A detailed study of the analytic properties of the q-integral representation of the $\mathcal{Q}$ operator will be discussed in future works. Also, it would be interesting to investigate on the relationships between the same q-integral representation and the Green's function of the Schr\"odinger equation corresponding to the interpolating flow of the B\"acklund transformations, on the line of the research developed in \cite{RZQ}.

\end{document}